\documentclass[12pt]{article}
\usepackage{graphicx,color}
\usepackage[usenames,dvipsnames]{pstricks}

\def\ben{\begin{equation}}
\def\een{\end{equation}}
\def\bea{\begin{eqnarray}}
\def\eea{\end{eqnarray}}
\def\vx{{\vec{x}}}

\def\vphi{{\vec{\phi}}}
\def\vpsi{{\vec{\psi}}}
\def\vk{{\vec{k}}}

\begin{document}
\begin{titlepage}
\thispagestyle{empty}
\begin{flushright}
UK/12-02
\end{flushright}

\bigskip

\begin{center}
\noindent{\Large \textbf
{Non-equilibrium Dynamics of 
$O(N)$ Nonlinear Sigma models: a Large-$N$ approach}}\\
\vspace{2cm} \noindent{
Sumit R. Das$^{a}$\footnote{e-mail:das@pa.uky.edu} and
Krishnendu Sengupta$^{b}$\footnote{e-mail:ksengupta1@gmail.com}}

\vspace{0.2cm}
  {\it
 $^{a}$Department of Physics and Astronomy, \\
 University of Kentucky, Lexington, KY 40506, USA \\
 $^{b}$ Theoretical Physics Department, \\Indian
Association for the Cultivation of Science, Jadavpur,
Kolkata-700032, India.\\
 }
\end{center}

\vspace{0.3cm}
\begin{abstract}

We study the time evolution of the mass gap of the $O(N)$ non-linear
sigma model in $2+1$ dimensions due to a time-dependent coupling in
the large-$N$ limit. 
Using the Schwinger-Keldysh
approach, we derive a set of equations at large $N$ which determine
the time dependent gap in terms of the coupling. These equations
lead to a criterion for the breakdown of adiabaticity for slow
variation of the coupling leading to a Kibble-Zurek scaling law.
We describe a self-consistent numerical procedure to solve these large-$N$
equations and provide explicit numerical
solutions for a coupling which starts deep in the
gapped phase at early times 
and approaches the zero temperature equilibrium 
critical point $g_c$ in a
linear fashion.
We demonstrate that for such a protocol
there is a value of the coupling $g=
g_c^{\rm dyn}> g_c$ where the gap function vanishes, possibly
indicating a dynamical instability. We study the dependence of
$g_c^{\rm dyn}$ on both the rate of change of the coupling and the
initial temperature. 
We also verify, by studying the evolution of the
mass gap subsequent to a sudden change in $g$, that the model does
not display thermalization within a finite time interval $t_0$ and
discuss the implications of this observation for its conjectured
gravitational dual as a higher spin theory in $AdS_4$.

\end{abstract}
\end{titlepage}
\newpage

\tableofcontents
\newpage

\section{Introduction and Summary}

The study of non-equilibrium dynamics of quantum field theories due to
a time dependent
mass or coupling has applications to many areas of physics. 
This has played a key role in our understanding
of quantum field theory in cosmological backgrounds \cite{Birrell:1982ix}
and has, in recent
years, been used as holographic descriptions of cosmological
solutions in asymtptotically anti-de-Sitter spacetimes
\cite{Das:2006dz}.
In condensed matter physics, the problem has
received a lot of attention in recent times due to its experimental
relevance to cold atom systems
\cite{sengupta,CCa,CCb,CCc,SCa,CCd,SCb,slowqc}. 
There are several theoretical motivations to study this problem. One
question relates to the issue of thermalization. Suppose we start with
a field theory in its ground state. Now consider changing the coupling
in time at some rate, eventually reaching some other constant value at
late times. The question is : does the system reach a steady state at
late times, and if so, does the steady state resemble a thermal state
in any sense ?

Another issue involves the dynamics of a field
theory when the time dependent coupling approaches or crosses an
equilibrium critical point. This problem is of relevance to, and was
initially studied in the context of, phase transitions in an expanding
universe \cite{KibbleZurek}.  In this case, adiabaticity will be
inevitably lost close to criticality and the subsequent dynamics will
carry universal signatures of the critical point. For slow dynamics, a
simple scaling hypothesis can indeed be used to show that several
quantities such as density of excitations and excess energy scales
with the rate of quench according some universal power-law which is
determined solely by the universality class of the critical point
\cite{sengupta}. Unlike equilibrium critical phenomena, there is no 
established theoretical framework to understand such universal behavior,
and for strongly coupled field theories there are few theoretical
tools to study this problem.
Remarkable exceptions in this regard are systems in $1+1$
dimensions, where methods of boundary conformal field theory can be
used to obtain results for correlation functions for an abrupt quench
from a massive theory to a critical
point\cite{CCa,CCb,CCc,SCa,CCd}. 
It is thus important to accumulate as many ``data
points'' as possible by examining individual models.

In this paper, we use large-N methods to study the non-equilibrium
dynamics of the $O(N)$ Nonlinear Sigma Model in $2+1$ dimensions. We consider a time
dependent coupling $g(t)= g_i+ (g_f-g_i)\tanh^2(vt)$ (where $v$ is the
ramp rate and $g_f$ and $g_i$ are initial and final values of $g$)
such that the system is initially, at $t=0$, in thermal equilibrium at
a temperature $T$ in the disordered (paramagnetic) phase, and study
the behavior of the mass gap of the model during and/or subsequent to
this dynamics. 

Using the Keldysh formalism, we show that at leading order in large
$N$ the gap function is determined in terms of the time dependent
coupling by a coupled set of differential and integral equations. For
an initially slow variation of the coupling, a derivative expansion
can be used to reduce these equations to a single {\em inhomogeneous}
differential equation for the gap function, where the departure of the
coupling from its equilibrium critical value acts as a source term.
Near the critical point, adiabaticity breaks down.  We determine the
condition for breakdown of adiabaticity and show that it leads to a
Kibble-Zurek type scaling relationship. We then describe a
self-consistent numerical procedure to solve the coupled set of
equations for arbitrary rates of change of coupling, and use this
procedure to study the time dependence of the gap function as we
approach close to the equilibrium critical point in the gapped phase.
We find that the gap function always vanishes at a time $t_0$ where
the instantaneous value of the coupling, $ g(t_0)=g_c^{\rm dyn}$ is
{\em larger} than the equilibrium critical point at $T=0$, $g_c^{\rm
  eq}$. (We use the normalization of \cite{sachdev1} where $g_c^{\rm
  eq} = 1$.)  We chart out the dependence of $g_c^{\rm dyn}$ on the
ramp rate $v$ and the initial temperature $T$.  We then study the
evolution of the mass gap of the model subsequent to a sudden change
in $g$ and verify that the system does not exhibit thermalization up
to a time $t_0$ till which we can numerically track such an
evolution. This result is consistent with known results about the lack
of thermalization of vector models in the large-N limit
\cite{largeNthermal} We comment on the consequence of such an absence
of "fast" thermalization (for $t\le t_0$) of the gravity dual of this
model. Finally, motivated by the derivative expansion, we study the
problem for a Landau-Ginsburg dynamics and compare the time evolution
with the $O(N)$ problem.

Large-$N$ quantum quench has been {\em indirectly} studied by using
the AdS/CFT correspondence \cite{Maldacena} to map large-$N$ limits of
strongly coupled field theories to classical gravity. In these
examples \cite{janik,dnt,bdas} (which involve $N \times N$ matrices,
rather than $N$ component vectors), it is almost impossible to solve
the field theory itself, but its dual gravity description is
tractable. In contrast, the dual formulation of the 2+1 dimensional
$O(N)$ vector model has been conjectured in \cite{Klebanov:2002ja} to
be a higher spin gauge theory in $AdS_4$ \cite{vasiliev} which
contains an infinite number of massless higher spin fields. This
conjecture has been explored in may papers, see in particular
\cite{Das:2003vw,yin} \footnote{For other holographic correspondences
  involving higher spin gauge theories, see \cite{sezgin}}.  In this
case the field theory is tractable, and it will be interesting to see
if this teaches us anything about quench and in particular
thermalization time of the higher spin theory, specifically because
there is an explicit dual map \cite{Das:2003vw}.

Quantum quench in the large-$N$ expansion has been studied earlier for
the {\em linear} sigma model in \cite{SCb} for infinitely fast
quenches. This work does not deal with the issue of scaling behavior
near the critical point.  A similar work deals with BCS theory with an
abruptly changing coupling \cite{barankov}. In contrast, our work in
the {\em nonlinear} model concentrates on the dependence of quench
dynamics on the rate of quench. The fact that $g_{dyn}$ is larger than
$g_c$ is similar in spirit to the phenomeon of stimulated
superconductivity found in \cite{elias} and studied in the AdS/CFT
context in \cite{eva}.

The plan of the rest of the paper is as follows. In Sec.\ \ref{mod1},
we use the Keldysh formalism to derive a set of equations which
determine the time-dependent gap (which we call the gap function) in
terms of the time dependent coupling.  This is followed by
Sec.\ \ref{slowdyn} where we derive the condition for breakdown of
adiabaticity for the model as we approach the equilibrium critical
point.  Next, we discuss the determination of $g_{dyn}$ of this model
in Sec.\ \ref{gdyn}. 
In Sec.\ \ref{gsud}  we discuss
the time evolution of the mass gap subsequent
to a sudden change of coupling.  Sec.\ \ref{conc} contains concluding
remarks. In the Appendix we study a Landau-Ginzburg dynamics
related to the $O(N)$ problem.

\section{The model and the gap equation}
\label{mod1}

The action of the model is given by
\ben
S = \int d^2x dt  \left[ \frac{N}{2 g(t)} (\partial_\mu \vphi (\vx,t))\cdot(\partial^\mu \vphi (\vx,t)) + \lambda
(\vx,t) [\vphi \cdot \vphi - 1] \right]
\label{1-1}
\een
where $\vphi (\vx,t)$ is a $N$ dimensional vector with real components. In the
large $N$ limit, $N \rightarrow \infty$ with $g(t)$ remaining $O(1)$.

The field $\lambda(\vx,t)$ is a Lagrange multiplier which imposes the
constraint $\vphi (\vx,t) \cdot \vphi (\vx,t) =1$. Redefining
fields
\ben
\vphi \rightarrow \vpsi = \sqrt{\frac{N}{g(t)}} \vphi
\een
the
lagrangian density becomes, up to a total derivative
\ben {\mathcal
L}= \frac{1}{2} (\partial \vpsi)^2 -\frac{1}{2} \Sigma (\vx,t) \left( \vpsi^2 -
\frac{N}{g(t)} \right) + N\frac{\alpha(t)}{g(t)}
\label{1-2}
\een
where
\ben \alpha (t) = \frac{1}{4} \left[ \frac{3}{2}\left(
\frac{\dot{g}}{g}
  \right)^2 - \left( \frac{\ddot{g}}{g} \right) \right],
~~~~~~~~\frac{1}{2} \Sigma(\vx,t) = \frac{g(t)}{N}\lambda(\vx,t) + \alpha (t)
\label{1-3}
\een
The last term in (\ref{1-2}) is field independent and can be therefore
ignored.

The partition function can be expressed in the Schwinger-Keldysh
formalism as \ben {\mathcal Z} = \int  {\mathcal D} \vpsi_{\pm}
{\mathcal D} \Sigma_{\pm}~e^{i \left[S(\vpsi_+,\Sigma_+) -
S(\vpsi_-,\Sigma_-)\right] } \label{keldysh} \een where we have
doubled all the fields as usual, and $S$ is the action for the
lagrangian in (\ref{1-2}). As is well known, the representation
(\ref{keldysh}) is schematic \cite{disorder1}- one has to pay
attention to the end point of the time contour. However these
"boundary" terms do not affect the saddle point equation, though
these are important for evaluation of the partition function by the
saddle point solution.

One can now integrate out the fields $\vpsi_\pm$
leading to the effective action for $\Sigma_\pm$,
\ben
S_{eff} =
N{\rm Tr}~\log (D^{-1}) - N\int d^2x dt \frac{1}{g(t)}[\Sigma_+ -
\Sigma_-] \een where $D$ is the propagator matrix whose inverse is
\ben
D^{-1} = \left(
\begin {array}{cc}
\partial^2-\Sigma_+&0\\
\noalign{\medskip}
0& -\partial^2 + \Sigma_-
\end {array}
\right) \een

The large-N saddle point equations therefore become \ben
\frac{1}{g(t)} = {\rm Tr}~D_{++},~~~~~~~\frac{1}{g(t)} = {\rm
  Tr}~(-D_{--})
\label{1-6}
\een
Therefore, the saddle has $\Sigma_+ = \Sigma_- \equiv \Sigma (t)$ and the equation
becomes
\ben
\frac{1}{g(t)} = \int d^2x ~ < \vx,t | D | \vx,t>_\beta
\label{1-7}
\een
where we are considering the problem at a temperature $T = 1/\beta$.
Note that the equality of  $\Sigma_+$ and $\Sigma_-$ is a feature of
the strict $N = \infty $ limit. Fluctuations around the saddle point
will destroy this equality.

The coincident Green's function in (\ref{1-7}) may be obtained by
considering a Heisenberg picture field which satisfies
the {\em homogeneous } equation
\ben
[ -\partial_t^2 + \partial_i^2 -\Sigma (t) ] \chi (\vx,t) = 0.
\een
Using a mode decomposition
\ben
\chi (\vx,t) = \int \frac{d^2k}{(2\pi)^2} [ a_k \chi_k(t) +
  a^\dagger_k \chi_k^\star(t) ]
\label{1-8}
\een
 the equal time Green's function $<\chi(\vx,t)
\chi(\vx^\prime,t)>_\beta$ is the two point function in
the thermal state, i.e. in the state
\ben
<a_k^\dagger a_k>_\beta = \frac{1}{e^{\beta \omega_0}-1}~~~~~~
<a_k a^\dagger_k>_\beta = 1 + \frac{1}{e^{\beta \omega_0}-1}
\label{1-9}
\een
where
\ben
\omega_0 \equiv \sqrt{\vk^2 + \Sigma(0)}
\label{1-10}
\een
The solution $\chi_k(t)$ may be written in the form
\ben
\chi_k(t) = \frac{1}{\sqrt{2\Omega_k(t)}}e^{-i\int^t
  \Omega_k(t^\prime) dt^\prime}
\label{1-11}
\een
where $\Omega_k(t)$ satisfies the equation
\ben
\frac{1}{2}\frac{\ddot {\Omega}_k}{\Omega_k} - \frac{3}{4} \left(
\frac{\dot{\Omega}_k}{\Omega_k} \right)^2 + \Omega_k^2 = k^2 +
\Sigma(t) \label{1-12}
\een
Then the gap equation (\ref{1-7})
becomes
\ben
\frac{1}{g(t)} = \int \frac{d^2k}{(2\pi)^2}
\frac{1}{2\Omega_k(t)} \coth (\frac{\beta \omega_0}{2}) \label{1-13}
\een
Note that the exponential factor in (\ref{1-11}) canceled in
the expression for the coincident time Green's function.

The equation (\ref{1-13}) has to be solved for $\Omega_k(t)$ for a
given $g(t)$ and substitution of the solution in (\ref{1-12}) gives
the gap function $\Sigma (t)$.

Before ending this section, we note that for a time independent
coupling $g_0$, our formalism reproduces the well-known equilibrium
solution \cite{sachdev1}. In this case, equation (\ref{1-12}) shows that $\Omega_k = k^2 + \Sigma$  i.e. the gap is
now independent of time. The integral
over $k$ on the right hand side yields the result \ben h_0 (\beta)
\equiv 2\pi \left( \frac{1}{g_c(\beta)} - \frac{1}{g_0} \right) =
\frac{2}{\beta} \log \left[ 2 \sinh (\frac{\beta}{2}
\sqrt{\Sigma_0}) \right] \label{1-14} \een Here $g_c(\beta)$ is
given by \ben \frac{2\pi}{g_c(\beta)} = \frac{2}{\beta} \log \left[
2 \sinh (\frac{\beta \Lambda}{2}) \right] \een and $\Lambda$ is the
momentum space UV cutoff. The equation (\ref{1-14}) can be solved
for $\Sigma_0$ \ben \sqrt{\Sigma_0} = \frac{2}{\beta} \log \left[
\frac{1}{2} e^{\beta
    h_0/2} + \frac{1}{2} \sqrt{e^{\beta h_0} + 4} \right]
\label{1-15} \een For any non-zero temperature this equation has a
solution. At exactly zero temperature the gap equation becomes \ben
h_0 = \sqrt{\Sigma_0} \een so that a real solution exists only for
$g_0 > g_c$. For $g_0 < g_c$ the $O(N)$ symmetry is spontaneously
broken and the theory is massless. The large-N solution presented
above is not valid in this phase,  though a valid solution in this
phase is well known \cite{sachdev1}.
 The point $g_0 = g_c$ is then a critical point which
separates the ordered and the massive phase.
At non-zero temperatures, the critical point becomes a cross-over
which separates regions of the phase diagram which are qualitatively
similar to the ordered and the disordered phases. The location of the
cross-over point is given by $\beta \sqrt{\Sigma_0} \sim 1$ which
implies $h_0 \sim T$ for small $T$.

\section{Quantum Quench : Breakdown of Adiabaticity}
\label{slowdyn}

When the coupling $g(t)$ varies slowly compared to the mass scale
set by the coupling itself, one expects that the gap function
$\Sigma (t)$ evolves adiabatically. Adiabaticity should break down
when  the gap function becomes small, e.g. near the zero temperature
critical point. Let us define \ben h(t) \equiv 2\pi (
\frac{1}{g_c(\beta)} - \frac{1}{g(t)} ) \label{1-17} \een To
investigate adiabaticity, we first need to find an expansion for
$\Omega_k(t)$ in terms of a $\Sigma (t)$ by solving (\ref{1-12}) in
a derivative expansion. This is easily done, and the lowest order
result is \ben \frac{1}{\Omega_k(t)} =
\frac{1}{\sqrt{k^2+\Sigma(t)}} +
\frac{\ddot{\Sigma}}{8(k^2+\Sigma(t))^{5/2}} - \frac{5
  \dot{\Sigma}^2}{32 (k^2+\Sigma(t))^{7/2}} + \cdots
\label{1-18}
\een
We then need to substitute this in (\ref{1-13}).

For zero temperature, it is possible to perform the necessary
integrals and the lowest order result is
\ben h(t) =
\sqrt{\Sigma(t)} - \frac{1}{24}\frac{\ddot{\Sigma}}{\Sigma^{3/2}} +
\frac{1}{32} \frac{\dot{\Sigma}^2}{\Sigma^{5/2}} + \cdots
\label{1-19}
\een
Inverting this (again in a derivative expansion)
we get \ben \sqrt{\Sigma (t)} = h(t) + \frac{1}{12 h(t)} \left(
\frac{\ddot{h}}{h} - \frac{1}{2}[\frac{\dot{h}}{h}]^2\right) +
\cdots \label{1-20} \een Therefore adiabaticity breaks down when
\ben | \frac{\ddot{h}}{h} - \frac{1}{2}[\frac{\dot{h}}{h}]^2 |  \sim
h^2 \een We will be interested in generic profiles of $g(t)$ for
which $h(t) \sim vt$ near $t=0$, e.g. $h(t) = - a \tanh(vt)$ for
some dimensional parameter $a$. For such a profile adiabaticity
breaks at a time \ben t_\star \sim v^{-1/2} \een
which is the usual
Kibble-Zurek scaling for linear quenches.

These result trivially extends to nonlinear quenches. We note that
the breakdown of adiabaticity can be also investigated analytically
for low temperatures and the results are qualitatively similar up to
exponentially small corrections.

\section{Numerical Results for a Dynamical Instability}
\label{gdyn}

In this section, we determine the value of the coupling where the gap
function first becomes zero.  This possibly signals a dynamical
instability. The protocol for $h(t)$ that we follow for studying this
phenomenon is the following: we start with a fixed $g=g_i$ inside the
disordered phase and with an equilibrium temperature $T$ and decrease
$g$ to $g_f < g_i$ at the end of the evolution with a speed $v$. In
the following we will take $g_i = 4$ in all the calculations.  This
protocol is realized using \ben g(t)= g_i +(g_f-g_i) \tanh^2(vt) \een
In what follows, we track the time evolution of the gap $\Sigma(t)$
and focus on finding the largest value of $g_f$ for which the minimum
value of the mass gap, $\Sigma_{\rm min}$, reaches zero at some point
during the evolution. This value yields $g_c^{\rm dyn}$. Since the
numerical solution of the gap equation becomes difficult when
$\Sigma_{\rm min} \to 0$, we extract the position of the dynamical
critical point by extrapolation of $\Sigma_{\rm min}$ as a function of
$g_f$. To elaborate, for each $v$, we let the system evolve from $g_i$
to $g_f$.  We vary $g_f$ and approach the static critical point till
it becomes difficult to obtain numerical convergence. We plot
$\Sigma_{min}$ as a function of $g_f$ for a given $v$ and extrapolate
this data to find $g_{dyn}$, {\it i.e.}, value of $g_f$ for which
$\Sigma_{min}$ reaches zero. We note at the outset that we have
checked that the curve fitting based on such extrapolation typically
generated correlation coefficient $R \ge 0.99$ and standard deviation
$\sigma \le 0.007$ which shows that errors from such a procedure are
minimal.
\begin{figure}[h!]
\begin{center}
\includegraphics[scale=1.0]{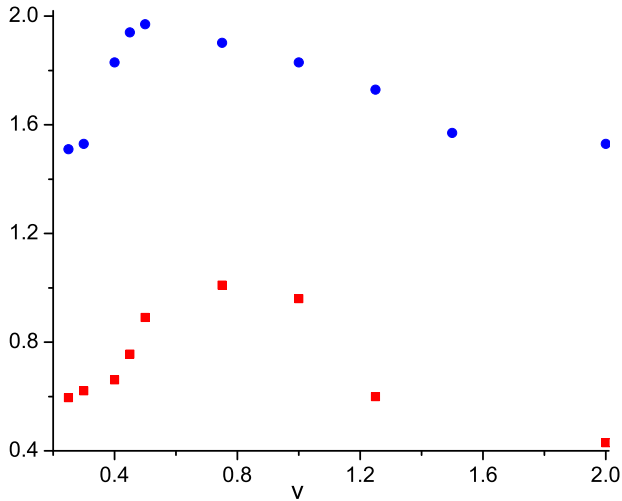}
\end{center}
 \caption{Rate dependence of $g_c^{\rm dyn}-1$ for $T=2$ (red square) and
$g_c^{\rm dyn}$ for $T=1.75$ (blue circle). The shift in the value
of $g_c^{\rm dyn}$ for $T=2$ is carried out to enhance clarity.}
\label{figdcp1}
\end{figure}

The numerical procedure we adopt for obtaining $\Sigma_{\rm min}$
for a given $v$ and $g_f$ is as follows. First, we provide initial
guess values of $\Sigma(t)$ for a discrete set of points $t=t_i$ and
numerically solve Eq.\ \ref{1-12} to obtain $\Omega(k_i,t_i)$ for an
array of discrete $k_i$ and $t_i$. From these values, using
interpolation, we compute the $k$ integral appearing in the right
side of Eq.\ \ref{1-13} and obtain a set of trial values $g_{\rm
trial}(t_i)$. We then minimize the function $[g(t_i)-g_{\rm
trial}(t_i)]^2$ self-consistently by varying $\Sigma(t_i)$. We note
here that the interpolation was carried out by choosing a set $N=21$
points. We have repeated some of our calculations with $N=31$ and
$N=41$ points and have found that the values of $\Sigma(t_i)$
obtained do not change to three decimal places. Thus the error bar
in the data from finite size of $N$ is minimal.

\begin{figure}[h!]
\begin{center}
\includegraphics[scale=1.4]{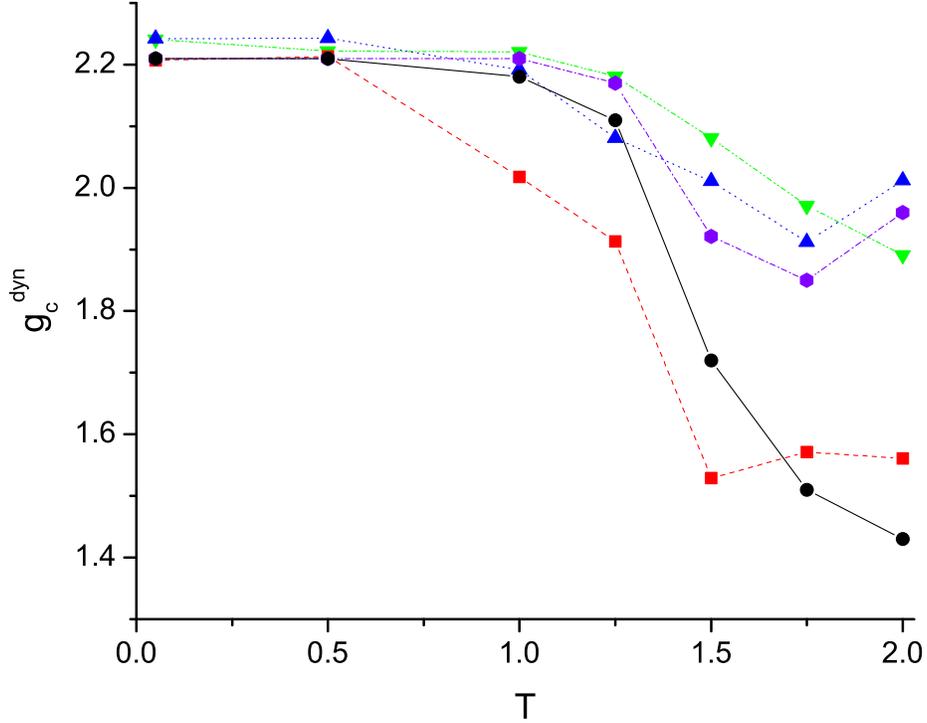}
\end{center}
 \caption{\label{figduncrit2}
Temperature dependence of $g_c^{\rm dyn}$ for $v=2$ (black circle),
$1.5$ (red square), $1.0$ (violet hexagon), $0.75$ (blue triangle),
and $0.5$ (green inverted triangle).} \label{figdcp2}
\end{figure}

The results of this procedure is summarized in Figs.\ \ref{figdcp1}
and \ref{figdcp2}. We find that for all rates $v\ge 0.25$ and
temperatures $T\ge 0.05$ that we have studied, the gap function
first touches zero at $g = g_c^{\rm dyn} > g_c=1$. The fact that
$g_{dyn}$ is larger than $g_c$ is similar in spirit to the
phenomenon of stimulated superconductivity found in \cite{elias} and
studied in the AdS/CFT context in \cite{eva}. We find from Fig.\
\ref{figdcp1} that the value of $g_c^{\rm dyn}$ increases with
increasing $v$, reaches a maximum around a critical rate $v^{\ast}$
which depends on the starting equilibrium temperature, and then
decreases as $v$ is further increased. We find $v^{\ast}$ reduces
with decreasing temperature and the hump flattens.

The existence of $v^{\ast}$ can be qualitatively understood as
follows. Since very slow dynamics in the disordered regime is
expected to be adiabatic, we expect $g_c^{\rm dyn}$ to approach
$g_c$ for small $v$. On increasing $v$, $g_c^{\rm dyn}$ increases
and deviates from $g_c$. This continues till a rate $v=v^{\ast}$
after which the system does not have enough time to respond to the
drive leading to a decrease in $g_c^{\rm dyn}$ with increasing $v$.
Note that here we have restricted our numerics to values of $v$ so
that the system reaches $g_f$ around $t=t_0$ till which we track the
dynamics. For faster $v$ and fixed $t_0$, the system will eventually
enter the quench regime where $g_c^{\rm dyn}$ will be determined by
the evolution of $\Sigma(t)$ with $g=g_f$ subsequent to the quench
till $t=t_0$. We do not address this regime in this section.

The temperature dependence of $g_c^{\rm dyn}$ for different $v$ is
shown in Fig.\ \ref{figdcp2}. Here we find that there is a crossover
regime around $T=1$ where the behavior of $g_c^{\rm dyn}$ changes
from the high temperature region where it shows appreciable
variation with $v$ to a low temperature regime where it becomes
virtually independent of $v$. Note that the saturation of $g_c^{\rm
dyn}$ for small $v$ is consistent with the shifting of $v^{\ast}$ to
lower values with decreasing temperature.

We end this section by noting that the dynamical instability of the
time-dependent mass gap studied above need not correspond to a phase
transition since the nature of the correlators of theory at the
point where the time dependent mass gap vanish need not be
long-range due to memory effects incorporated in $\Omega_k(t)$. It
would be interesting to study such correlators near the instability;
presently the slow convergence of the numerical solution near the
instability prevents us from carrying out such a detailed study.

\section{Time evolution after a sudden quench}
\label{gsud}

\begin{figure}[h!]
\begin{center}
\includegraphics[scale=1.0]{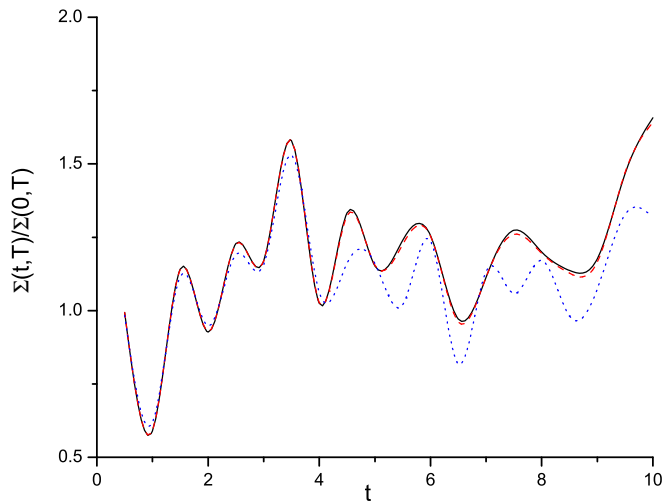}
\end{center}
 \caption{\label{figimp}
The evolution of the mass gap after a sudden quench at $t=1$ as a
function of time till $t=t_0=10$ for temperatures $T=2$ (blue dotted
line), $T=1$ (red dashed line) and $T=0.1$ (black solid line).}
\end{figure}

In this section. we discuss the time evolution of the mass gap
$\Sigma(t)$ subsequent to a sudden impulse imparted to the system.
The impulse is imparted by changing $g(t)$ as $g(t)= (g_i-g_d) + g_d
\tanh^2[v(t-t_1)]$ with $g_i=4$, $g_d=1$, $t_1=1$ and $v=20$. Note
that with this choice, the system starts its evolution with $g=g_i$
at $t=0$. Near $t=t_1$, the coupling $g$ changes to $g_f=g_i-g_d$
and back to $g_i$. The change takes place within a time window of
$\tau \sim 1/v$ around $t_1$ and thus appear as an instantaneous
impulse for large $v$. The plot of the subsequent evolution of
$\Sigma$ is plotted as a function of time in Fig.\ \ref{figimp} for
several initial temperatures. We note that for all temperatures, the
system does not show any sign of thermalization in the sense that
$\Sigma(t)$ does not approach any constant steady-state values till
the time $t_0=10$ that we track it's evolution numerically.

We believe that this is a manifestation of the lack of
thermalization in $O(N)$ vector models to leading order in $1/N$. We
note that from our numerical result, we can not rule out
thermalization of the system at longer times; however, the model
certainly do not thermalize for $t\le t_0$.  This is consistent with
the expectation that vector models do not thermalize at large $N$,
which is due to the lack of quasiparticle scattering at $O(1/N)$. (
Such scattering in the present models appears in $O(1/N^2)$ and
ultimately leads to thermalization.) This is in contrast with
large-N models of matrices, where thermalization is expected to
occur \cite{largeNthermal} at the leading order \footnote{The lack
of thermalization is consistent with the results of
\cite{Buividovich}.}. This is manifest for the class of such large-N
models which have gravity duals, where thermalization is seen as
formation of black holes \cite{janik}. In such models,
thermalization is almost instantaneous for local operators.

This is also consistent with the fact that the gravity dual of the
$O(N)$ model is a higher spin gauge theory rather than standard
Einstein gravity.  In usual $AdS/CFT$ duals of models of large-N
matrices (e.g. gauge theories), thermalization is signalled by black
hole formation. However, a study of the finite temperature properties
of the singlet sector of the higher spin model shows that there is no
large-N transition at order one temperatures \cite{shenker}. This
possibly implies the absence of thermodynamically stable large black
holes with order one Hawking temperatures in this higher spin theory.

\section{Conclusions}
\label{conc}

In conclusion, we have shown that the nonequlibirum dynamics of the
non-linear O(N) sigma model with a time dependent coupling $g(t)$ is
summarized by two coupled equations (\ref{1-12}) and
(\ref{1-13}). Whenever a derivative expansion is valid, these
equations can be reduced {\em formally} to a single differential equation for the gap
function with arbitrarily higher derivatives, equation (23).
This latter equation has been used to determine the
condition of breakdown of adiabaticity at zero temperature : this
led to Kibble-Zurek scaling.  We then described how these equations
can be solved by a self-consistent numerical procedure.  We presented
numerical results for the case where the coupling which approaches a
constant value deep in the paramagnetic region, and approaches the
zero temperature equilibrium critical value in the future, starting
with a thermal state.  We found that during the time evolution, the
instantaneous value of the gap reaches zero for a coupling $g_{dyn}$
which is {\em larger} than the coupling at the equilibrium critical
point and studied the dependence of $g_{dyn}$ on the quench rate.
Finally, we have studied the response of the model to a sudden quench
and verified that the model does not exhibit short-time
thermalization, and have discussed the consequence of this phenomenon
for its gravity dual.

Our numerical procedure does not work well near $g=g_{dyn}$ which is
the regime of obvious interest.  We are working on different numerical
procedures to overcome these difficulties.

It would be interesting to extend this approach to discuss the
dynamics when the coupling {\em crosses} the critical point. This
requires a treatment of the saddle point equations in the ordered
phase as in \cite{sachdev1}. Finally, our work can in principle be
extended to a study of periodic dynamics. These questions are being
currently investigated.

\section{Acknowledgements}
\label{ack}

We are grateful to Ganpathy Murthy for extensive discussions and
collaboration at the early stages of this work and for valuable
comments about the manuscript. S.R.D. would like to thank  Institut
de Fisica Teorica at Madrid, Tata Institute of Fundamental Research
at Mumbai, Indian Association for the Cultivation of Science at
Kolkata and Kobayashi-Maskawa Institute in Nagoya for hospitality
during the course of this work. This work is partially supported by
National Science Foundation grants PHY-0970069 and PHY-0855614. KS
thanks University of Kentucky for hospitality and DST, India for
support through grant no. SR/S2/CMP-001/2009.

\section{Appendix : Landau-Ginzburg Dynamics}
\label{LG}

Large N theories are classical in the leading order, and one might
imagine that there is a classical equation of motion which describes
this limit, typically in one higher dimension.  Models of $N \times N$
matrices are often of this type, e.g. matrix quantum mechanics whose
large-N limit is described by the classical equations of $1+1$
dimensional string theory \cite{brezin} - the string field in this
case is in fact a single massless scalar which can be identified with
a suitable collective variable \cite{c=1sft}. AdS/CFT dualities are
also of this kind : the large-N classical theories are generically
{\em string} theories in one higher dimension and contain an infinite
number of fields. Nevertheless, usually in the strong coupling limit,
only a few of these fields are massless (which is the statement that
in the field theory only a few operators have dimensions of order one
rather than of order $N$). In that case the dual theory is described
by a finite number of classical equations of motion, viz. Einstein
equations and equations of motion of a few other fields.

So long as the derivative expansion is valid, the gap function in our
system satisfies an {\em inhomogeneous} differential equation
(\ref{1-19}) where the quantity $h(t)$ defined above acts as a
source. This equation is valid in the disordered phase. In fact, in
the general case where the lagrange multiplier field $\Sigma$ depends
on both space and time, this is one of the collective fields which
describe the large-N dynamics. Exactly at the critical point this is
identified with the scalar field in $AdS_4$ Vasiliev theory
\cite{Giombi:2011ya}. Of course the equation (\ref{1-19}) breaks down
as we approach $h(t) = 0$.

This motivates us to compare the nonequilibrium behavior of our
$O(N)$ model with the non-equilibrium dynamics of a toy
Landau-Ginzburg (LG) type model. The equation is given by \ben
\frac{d^2 \xi (t)}{dt^2} +c \frac{d\xi}{dt} + \xi^3 +J(t) = 0
\label{2-2} \een 
The field $\xi$ is the analog of the gap function
and $J(t)$ plays the role of the coupling $h(t)$ in the previous
sections. A "mass term" in this model has been set to zero to ensure
that the equilibrium theory is critical when $J(t)=0$ and we have
added a friction term with coefficient $c$ to account for dissipation.

\begin{figure}[h!]
\begin{center}
\includegraphics[scale=0.75]{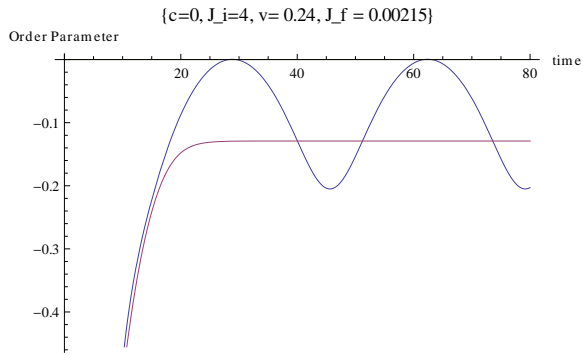}
\end{center}
 \caption{\label{fig:five}
A typical behavior of the order parameter as a function of time, with the source function
given by (\ref{2-2s}) and vanishing friction.
The evolution starts from $t=0$. The purple
line is the adiabatic value of the order parameter. The value of $J_f$
 chosen in this plot corresponds to the dynamical critical point.}
\end{figure}

\begin{figure}[h!]
\begin{center}
\includegraphics[scale=0.75]{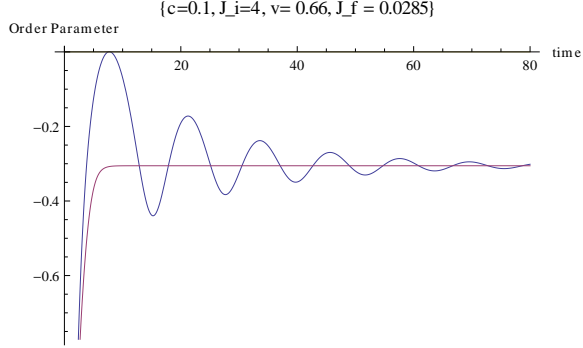}
\end{center}
 \caption{\label{fig:five2}
A typical behavior of the order parameter as a function of time, with the source function
given by (\ref{2-2s}) and some friction.
The evolution starts from $t=0$. The purple
line is the adiabatic value of the order parameter. The value of $J_f$
 chosen in this plot corresponds to the dynamical critical point.}
\end{figure}

Consider a
time dependent source in (\ref{2-2}) of the form \ben J(t) = J_i +
(J_f - J_i) \tanh^2(vt) \label{2-2s} \een pretty much like $g(t)$
chosen for the $O(N)$ model. We now solve the equation for a fixed
$J_i$, starting at $t=0$ with adiabatic initial conditions for
different values of $J_f$, and determine the value of $J_f =
J_f^{c}$ for which the order parameter $\xi$ first touches zero. A
typical time evolution of the order parameter is shown in
Figure(\ref{fig:five}) for vanishing friction and in
Figure(\ref{fig:five2}) in the presence of some friction.

\begin{figure}[h!]
\begin{center}
\includegraphics[scale=0.75]{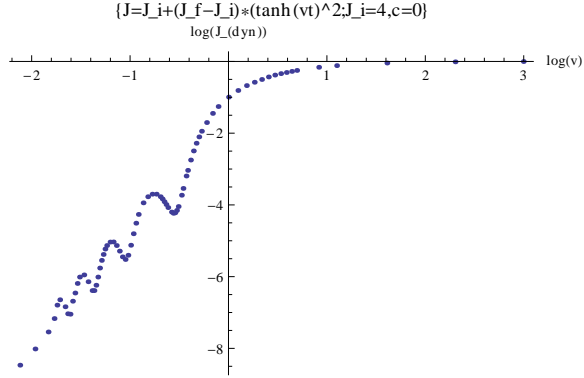}
\end{center}
 \caption{\label{fig:six}
A log-log plot of $J_{dyn}$ versus $v$ with no friction. The saturation
value is $\frac{1}{4}J_i$ as expected.}
\end{figure}

\begin{figure}[h]
\begin{center}
\includegraphics[scale=0.75]{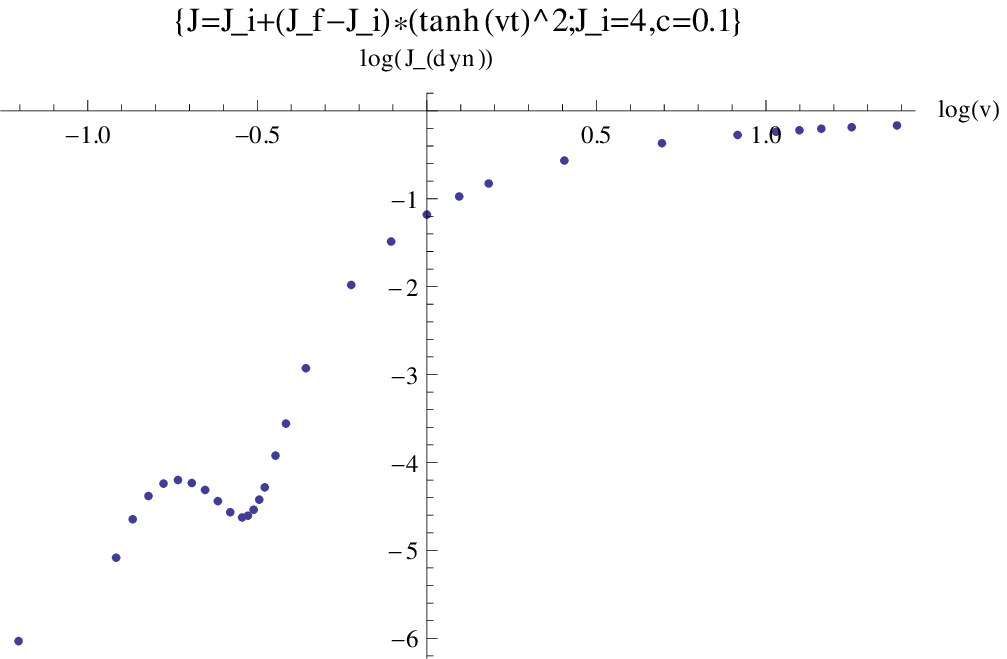}
\end{center}
 \caption{\label{fig:six1}
A log-log plot of $J_{dyn}$ versus $v$ with friction. }
\end{figure}

Figure (\ref{fig:six}) shows the behavior of $J_{dyn}$ as a function
of the ramp speed for vanishing friction. Figure (\ref{fig:six1}) is
the same plot in the presence of friction.
For very small $v$, $J_{dyn}$ is very small and
increasing with $v$ since the time evolution is expected to be
adiabatic. For large $v$ we have a rapid quench. The system does not
have enough time to react to the change and remains in the initial
state for quite some time, after which it starts oscillating. $J_{dyn}$
saturates to a constant value.

For vanishing friction, the saturation value may be
understood as follows. For large $v$, the quench is rapid at $t=0$.
We may then approximate $J(t)$ by $J(0) = J_i$ and $J(t) = J_f$ for
$t > 0$ .  Then a first integral of the equation of motion
(\ref{2-2}) for $t > 0$ is given by \ben \frac{1}{2}
\left(\frac{d\xi}{dt} \right)^2 + \frac{1}{4} \xi^4 +  J_f \xi = E
\een Since at $t=0$ the initial conditions are adiabatic $\xi(0) =
[-J_i]^{1/3}$ and ${\dot{\xi}}(0) = 0$, so that $E=
\frac{1}{4}J_i^{4/3} - J_f J_i^{1/3}$. Now, at the time when the
order parameter first touches zero, we must have $\xi =0$ and
${\dot{\xi}}=0$. For this to happen we must have $E=0$, i.e. $J_f =
\frac{1}{4} J_i$. Indeed, we have checked that the saturation value
of $J_{dyn}$ is indeed given by $\frac{1}{4} J_i$.

The behavior of $J_{dyn}$ as a function of $v$, however, displays
non-monotonic behavior for intermediate $v$, with multiple
well-defined humps. In the presence of friction we could not find
multiple humps, but one hump always remains. We do not understand the
reason behind this behavior. However, it is interesting that there is
a similar behavior in the nonlinear sigma model, even though the
latter differs significantly from our toy model when the gap vanishes
at some time.

The presence of humps in the behavior of $J_{dyn}$ is intruigingly
similar to similar non-monotonicity found in the $O(N)$
model. However, other aspects of these results appear to be quite
different from our $O(N)$ results. In particular, while $J_{dyn}$ 
in the LG theory saturates at large speeds, the $g_c^{dyn}$
appears to first rise and then come closer to the equilibrium critical
coupling $g_c=1$. It is conceivable that for much larger speeds,
$g_c^{dyn}$ rises again and saturates. However, numerical convergence
becomes difficult at high ramp speeds.

It is also conceivable that this difference reflects a fundamental
difference between the large-N classical theory of $O(N)$ as
formulated above and a LG type theory. As discussed above the gap
function is not related to the coupling in a local fashion in
time. The gap equation (\ref{1-13}) relates $g(t)$ with $\Omega_k(t)$
which depends on the entire function $\Sigma (t)$ through the
differential equation (\ref{1-12}). It is only in the adiabatic
approximation that the gap function satisfies a differential equation
with a source given by the coupling - and adiabaticity of course fails
when the gap function vanishes. This is related to the fact that
this theory is dual to a theory of an {\em infinite number} of
massless higher spin
fields, and the gap function is simply one of these fields. This implies
that an effective equation for the gap function would be
non-local. It would be interesting to see if the bilocal collective
field theory approach to this duality \cite{Das:2003vw} can shed any
light on this issue.

\end{document}